\documentstyle[preprint,aps]{revtex}

\begin{document}
\draft

\title{Quantitative Characterization of Permeability Fluctuations in Sandstone}

\author{Hern\'an A. Makse,$^1$ Glenn W. Davies, $^2$
Shlomo Havlin$^{1,3}$, Plamen Ch. Ivanov,$^1$ Peter R. King,$^2$
{}~and~H.~Eugene~Stanley$^1$}

\address{$^1$Center for Polymer Studies and Dept. of Physics, Boston
University, Boston, MA 02215 USA\\
$^2$BP Exploration Operating Company Ltd.,
Sunbury-on-Thames, Middx., TW16 7LN, UK\\
$^3$Department of Physics, Bar-Ilan University, Ramat Gan, ISRAEL}

\date{\today}

\maketitle

\begin{abstract}
Sedimentary rocks have complicated permeability fluctuations arising
from the geological processes that formed them.  These permeability
fluctuations significantly affect the flow of fluids through the rocks.
We analyze data on two sandstone samples from different geological
environments, and find that the permeability fluctuations display
long-range power-law correlations characterized by an exponent $H$.  For
both samples, we find $H \approx 0.82-0.90$.

\end{abstract}


\narrowtext

\section{INTRODUCTION}

Permeability in sandstone can change by many orders of magnitude over
very short distances \cite{allen}.  Not only are there large
fluctuations in permeability but the permeability can exhibit strong
anisotropy.  Deriving a mathematical representation to describe these
spatial fluctuations in permeable rocks is a major challenge.  This
represents an important technological problem as hydrocarbon is
frequently found in such rocks and its efficient recovery depends on
the ability to describe and predict flow through such rocks, as does
contaminant dispersal and control in groundwater.

Previous studies have concentrated on spatial fluctuations in porosity
\cite{hewett}.
For clean, well sorted sands porosity is
independent of grain size while the permeability is known to be
proportional to the square of the particle radius.  For poorly sorted,
non-clean sands permeability is usually not a unique function of
porosity\cite{sahimi,ho}.
Also permeability directly affects fluid flow so it is more
relevant when studying hydrocarbon recovery and groundwater flow.

Permeability fluctuations have been traditionally modeled assuming
correlations of finite range \cite{porosityshort}.
Long-range correlations in permeability
significantly affect the large-scale flow in a porous
medium,  so it is of practical importance to distinguish between
long-range and short-range correlations.
In this work, we analyze
detailed permeability maps, and find that the data display long-range
correlations.  Specifically, we investigate two different sandstone
samples.
One is a Triassic, {\it fluvial} trough cross bedded sandstone from
Hollington near Stafford in the East Midlands of England
\cite{ho}.  The second sample is a Triassic, planar-tabular cross
bedded
{\it aeolian} sandstone from Locharbriggs near Dumfries, Scotland.
Although the geological process is different for both samples, in both
cases size segregation of particles occurs by avalanches---modified by
fluid flow in one case.  We find the exponent characterizing the
long-range correlations in the permeability values are the same,
within  errors bars, for both
types of sandstone: aeolian and fluvial.

\section{THE SAMPLES}

The Hollington sample (Sample Ho) is an example of trough cross
bedding (Fig. \ref{Ho}).
Trough cross bedding results from the
migration of sand bars (accumulations of sand) along a river bed
\cite{allen}.  The current
progressively moves sand from the upstream side of the sand accumulation to the
downstream.
On reaching the downstream side the sand will roll down this face
once a
critical angle is exceeded. There is a degree of turbulence or eddying
at the
downstream face. This results from higher flow rate at the top of the
face (foresets) and lower at the base (bottomsets).
Hence, grains towards the top
of the downstream slope are better sorted and have lower packing. Those at the
base are more poorly sorted,
resulting from a mixture of larger grains rolling
down the lee face and
finer material dropping out of suspension. The degree
of permeability contrast between
the top and base of the bed depends on a
combination of stream velocity,
turbulence and available grain sizes. If
velocity is very high and turbulence
low foresets will be formed continuously
from material avalanching down
the downstream face, with little contrast
between lamina thickness,
grain size or sorting. If turbulence and velocity
fluctuate more the thickness,
grain size and sorting of individual laminae
will vary.

In Fig. \ref{Ho} can be seen a sharp
boundary between three regions in the sample.
These represent where one
sand bar has been overridden by another
partially eroding the top of the
lower unit. This has
occurred twice. Two of the zones are
similar in character and have higher
permeability whereas the third
has consistently lower permeability. We,
therefore, treat the sample as
having one zone of ``low'' and two of ``high'' permeability.

The Locharbriggs sample (Sample Lo) is shown in Fig. \ref{Lo}.
The sedimentary rock was formed by windblown sand \cite{bagnold}.
The wind
transports the sand to the crest of the dune. When the sand exceeds a
critical angle it avalanches. In these avalanches the fine and coarse
grains systematically segregate. The resulting layers are clearly
seen in Fig.  \ref{Lo}. In this case the eddying fluid (air) on the leeward
face has insufficient energy to influence the avalanching process
resulting in the foresets being planar rather than convex.
Note
that in this system grain size range and permeability is relatively
constant parallel to the laminations. This mechanism of grain size
segregation is explained in more detail in Ref. \cite{makse}.
Again merging of dunes gives rise to sharp transitions in
permeability,
hence there are two zones. We notice also a
strong anisotropy in the high permeability zone of
this sample.

\section{CORRELATIONS}

The permeability of the rock of both samples
was measured by standard mini-permeametry \cite{permeametry}.
The mini-permeameter
essentially measures the
pressure required to force a given flow rate of gas into the rock.
The permeability maps are shown in Figs. \ref{Ho}b and \ref{Lo}b.
It is clear that permeability varies significantly
within a very short scale.
The permeability may not be an independent random process. To
measure the size of the fluctuations we plot the permeability
histograms
for the Sample Ho in Fig. \ref{distributionHo}.
The distributions for the high and low permeability regions
are separated. The high permeability zone has typical
permeability 3300mD, the low permeability zone has
typical permeability 30mD. It should be noted that for uncompacted,
well sorted, clean, quartzite sandstone,
local permeability is proportional to the
square of the grain radius. The high permeability zone consists of
interbedded fine and coarse grain material and hence has a much higher
variability. The low permeability zone is more homogeneous, consisting
of more exclusively fine grained material.

Next the spatial correlations in permeability were measured. We study
the correlations of
the permeability field  $k(i,j)$ ($i,j=1,...,n_x, n_y$) along the
$x$ and $y$  directions (see Fig. \ref{Ho}b).
To this end, we first integrate the permeability
variables along both directions separately, by
calculating the
``net displacements''   $x_j(\ell)$  and $y_i(\ell)$ \cite{saupe}
\begin{mathletters}
\label{x}
\begin{equation}
x_j(\ell) \equiv \sum_{i=1}^{\ell} ~ ( k(i,j)-\overline{k(j)})
\qquad [j=1,...,n_y],
\end{equation}
\begin{equation}
y_i(\ell) \equiv \sum_{j=1}^{\ell} ~ (k(i,j)-\overline{k(i)}) \qquad
[i=1,...,n_x],
\end{equation}
\end{mathletters}
where $\overline{k(j)} = (1/n_x) \sum_{i=1}^{n_x} k(i,j)$ and
$\overline{k(i)} = (1/n_y) \sum_{j=1}^{n_y} k(i,j)$.
Then we calculate the variance
$V_x(\ell)\equiv\langle \overline{x(\ell)^2} -
\overline{x(\ell)}^2 \rangle ^{1/2}$ and  $V_y(\ell)\equiv\langle
\overline{y(\ell)^2}-\overline{y(\ell)}^2 \rangle ^{1/2}$
as a function of the lag $\ell$.
The
spatial average over a window of size $\ell$
is denoted by the overbar, and the disorder average over
different displacements ($x_j$ and $y_i$)
is denoted by the angular brackets. The scaling behavior
of the variance  $V(\ell) \sim \ell^H$
can distinguish between short and long range correlations. For
uncorrelated permeability variables,
$H=1/2$, while $1/2<H<1$ indicates persistent
long-range correlations among the variables. The correlation
exponent $H$  describes
the ``roughness of the permeability landscape'' \cite{saupe,fbm}.
The method described so far is the
standard rms fluctuations analysis which however,
is known to fail in the
following situations: {\it (i)}
when the signal is nonstationary \cite{ck}, and {\it (ii)} when the
signal is highly correlated $H\simeq 1$ \cite{lesh}.
In
case {\it (i)},
the method detects spurious correlations due to the patchiness of
the signal\cite{ck},
while in case {\it (ii)} the method gives smaller effective
exponents (in particular when small samples are used) because the
variance has an upper bound $V(\ell) < \ell$ and therefore the method
cannot detect fluctuations with exponent $H \ge 1$ \cite{lesh}.
Apart from possible nonstationarities,
in
our case we find that the permeability values are strongly correlated.
To overcome the limitations of the rms method, we will analyze
the spatial
correlations of the permeability by
 using the detrended fluctuation analysis (DFA) of Ref.
\cite{ck} and the wavelet analysis \cite{wavelet}.

The results for the permeability correlations
for the Sample Ho are shown in Fig.  \ref{varianceHo}.
In Fig.  \ref{varianceHo}a we show the correlations
for both high and low
permeability zones combined measured in the  $x$ and $y$  directions.
In this case, before calculating the variance, the permeability is
normalized by dividing by the standard
deviation calculated independently for each direction.
The
data are consistent with  power-law correlations;
using the DFA method, we find
$H_x = 0.89\pm0.06$ and $H_y = 0.90\pm0.06$.
Results for the correlations along the $x$ direction
are shown separately for the
high and low permeability zones in Fig.  \ref{varianceHo}b.
The
correlations are satisfactorily
modeled by a power law where $H_x\simeq 0.89$,
independent of the magnitude of the overall permeability.
These values
are confirmed, within the error bars,
 using the wavelet analysis. We find that $H_x=0.82\pm0.06$
and $H_y=0.84\pm0.06$.

As seen in Fig. \ref{Lo} the high permeability zone of sample Lo
presents strong
anisotropy with anisotropic axes  $(x', y')$ not coincident
with the coordinate frame  $(x, y)$ (see Fig. \ref{Lo}b).
We calculate the variance along the $y'$ direction (parallel to the
direction of the crests) and find using the DFA method
$H=0.85\pm0.06$ (Fig. \ref{varianceLo}); a
value that is consistent with the findings for the Sample Ho. Using
wavelet analysis, we find $H = 0.84\pm0.06$.
Along the $x'$ direction a periodic morphology is observed
with a wave length of about 60 mm. This introduces a characteristic
length scale so that  no scale invariance power law correlations are
expected along this direction. The existence of this laminar periodic
structure is consistent with a depositional model of sand dune
dynamics
\cite{makse}.

\section{DISCUSSIONS}

Spatial fluctuations in rock permeability
exist, and require quantitative
methods to describe them. We have shown that
permeability correlations in two different rock samples
can be well described using a power law. Further, the
exponent is quite similar for the two samples.
The essential physical feature in
the formation of these two samples is avalanching of the sand grains,
which gives rise to segregation and alternation in the fine and coarse
material. It is possible that this mechanism
could explain the apparent
universality in the power law correlation for these rock types.

These spatial fluctuations have significant consequencies for
prediction of,
e.q., hydrocarbon recovery or contaminant transport in
groundwater \cite{sahimi,sona}.
The fact that there exist long range correlations implies
that the spread in contaminant transport could occur
be much faster than would
be predicted from a short range correlation model.

The authors would like to thank
BP Exploration Operating Company for financial support and
permission to publish this paper.
H. A. M. wish to thank S. Tomassone for valuable discussions.
S. H. wish to thank the Israel
Science Foundation for financial support.

\begin{figure}
\narrowtext
\caption{$(a)$ Photograph
of one of the slabs for
the Sample Ho. The sample consist of two slabs of  $474$ mm by
$276$ mm  and $10$ mm thick. Three faces at height $z=0$,
$z=10$, and $z=20$ mm  were used to study
the permeability pattern, from which the $z=0$ face is shown in this
figure. Unfortunately the measurements of one face were
corrupted by instrumentation
error and so only three faces could be used.
 $(b)$ Permeability map
of 1a.
The permeability was measured
every $10$mm in the $x$ direction and every $4$mm in the $y$ direction.
Therefore, a grid of $n_x=48$ by $n_y=69$ permeability values
was obtained.}
\label{Ho}
\end{figure}

\begin{figure}
\narrowtext
\caption{$(a)$ Photograph of the face  $z=0$
of the Sample Lo.
The sample consist of two slabs of $448$mm by
$246$mm  and $10$ mm thick, and again only three faces were used in
this study.
$(b)$ Permeability map
of 2a.
The permeability was measured every
$12$mm and $4$mm in the $x$ and $y$ directions, respectively, so that
a grid of $n_x=38$ by $n_y=61$ was obtained.
Notice the strong anisotropy of this
sample manifested by the crests elongated along the $y'$ direction.}
\label{Lo}
\end{figure}

\begin{figure}
\narrowtext
\caption{Normalized permeability distributions for the Sample Ho
corresponding to $(a)$ low permeability zone, and $(b)$ high
permeability zone. In both figures we plot the
distributions corresponding
to three different faces of the sample. The
distributions are fitted by Gaussian forms. We notice the
large difference in the mean value of the permeability between the low and
high permeability  zones.}
\label{distributionHo}
\end{figure}

\begin{figure}
\narrowtext
\caption{Log-log plot of the variances  of the permeability calculated
using the DFA method.
 $(a)$
Variances $V_x(\ell)$ and $V_y(\ell)$
 along the $x$ and $y$ directions respectively
averaged over the  three different faces of the sample, and
over the high and low permeability zones together for $V_x(\ell)$ and over the
high permeability zone for  $V_y(\ell)$.
The power law
relationship between the variance and the separation distance $\ell$
is characterized by  exponents $H_x=0.89\pm0.06$ and
$H_y=0.90\pm0.06$.
The
exponents are the same within error bars indicating the isotropy of the
correlations in the $x y$ plane.
$(b)$
Variance $V_x(\ell)$
calculated along the $x$ direction for the high
and low permeability zones, separately. Data are averaged over the three
different faces of the sample. Both set of data are
consistent with a
power law $H_x\simeq0.89$, showing that the spatial correlations are the
same in both zones.}
\label{varianceHo}
\end{figure}

\begin{figure}
\narrowtext
\caption{Log-log plot of the variance
calculated along the $y'$
for the high permeability zone of Sample Lo,
averaged over the three
different faces of the sample.
Along the $y'$ direction, a correlation exponent of $H=0.85\pm0.06$
is found. However, along the $x'$ direction,
a periodic pattern is observed.
Thus the anisotropy in
this sample is manifested in a change of behavior from long-range
correlation scaling along $y'$ to periodic morphology along $x'$.}
\label{varianceLo}
\end{figure}



\begin{references}
\bibitem{fbm}
The exponent $H$ is called the Hurst exponent for the associated
fractional Brownian motion (fBm) defined by Eq. (\ref{x}). However, it
should be noted that the permeability is a fractional Gaussian
noise (fGn).
\end{references}
\end{document}